\documentclass[letterpaper,twocolumn,prl,
aps,showpacs,groupedaddress,floatfix]{revtex4-1}
\usepackage[latin1]{inputenc}
\usepackage{bm}
\usepackage[usenames]{color}
\usepackage{multirow}
\usepackage{amssymb}
\usepackage{amsbsy}
\usepackage{mathtools}
\usepackage{amsmath}
\usepackage{stmaryrd}
\usepackage{graphicx}
\usepackage{epsfig}
\usepackage{placeins}
\usepackage{ulem}
\usepackage{bbold}
\usepackage{braket}
\usepackage{blindtext}
\usepackage[colorlinks,linkcolor=blue,citecolor=blue,urlcolor=blue]{hyperref}

\makeatletter

\begin{document}
\title{Simulating hydrodynamics on noisy intermediate-scale quantum devices 
with 
random circuits}

\author{Jonas Richter}
\email{j.richter@ucl.ac.uk}
\affiliation{Department of Physics and Astronomy, University College London, 
Gower Street, London WC1E 6BT, UK}

\author{Arijeet Pal}
\affiliation{Department of Physics and Astronomy, University College London, 
Gower Street, London WC1E 6BT, UK}

\date{\today}

%---------------------------------------------------------------------------------------------------
\begin{abstract}

In a recent milestone experiment,
Google's processor Sycamore heralded the era of 
``quantum supremacy'' by 
sampling from the output of \text{(pseudo-)}random circuits. We show that such 
random circuits provide tailor-made building blocks for 
simulating quantum many-body systems on noisy intermediate-scale quantum (NISQ) 
devices. Specifically, we propose an algorithm consisting of a random circuit 
followed by a trotterized Hamiltonian time evolution to study 
hydrodynamics and to extract transport coefficients in the linear response 
regime. We numerically 
demonstrate the algorithm by simulating the buildup of spatiotemporal 
correlation functions in 
one- and two-dimensional quantum spin systems, where we particularly 
scrutinize the inevitable impact of errors present in any realistic 
implementation. 
Importantly, we find that the hydrodynamic
scaling of the correlations is highly robust with respect 
to the size of the Trotter step, which opens the door to reach 
nontrivial time scales with a small number of gates. While 
errors within the random circuit are shown to be irrelevant, we furthermore 
unveil that meaningful results can be obtained 
for noisy time evolutions with error rates achievable on near-term hardware. 
Our work emphasizes the practical relevance of random circuits on NISQ devices 
beyond the abstract sampling task.
  
\end{abstract}

\maketitle
%-------------------------------------------------------------------------------
%--------------------

%
\begin{figure}[tb]
 \centering
 \includegraphics[width=0.85\columnwidth]{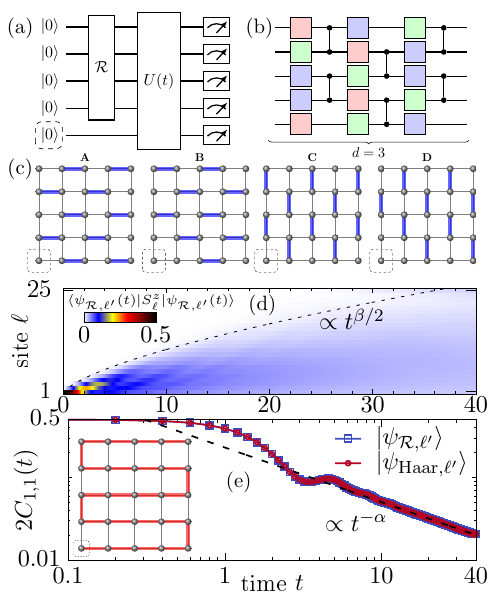}
 \caption{(a) A random circuit ${\cal R}$ acts on $L-1$ qubits, 
followed by a time evolution $U(t)$ on all 
$L$ sites. [(b),(c)] ${\cal R}$ comprises $d$ cycles, each composed  
of layers of one- and two-qubit gates. We consider a 2D geometry and ${\bf A} - 
{\bf D}$ are patterns of two-qubit gates used in 
different cycles. [(d),(e)] For reference site $\ell' = 1$, $\bra{\psi_{{\cal 
R},\ell'}(t)} S_\ell^z \ket{\psi_{{\cal R},\ell'}(t)}$ yields the correlation 
function 
$2C_{\ell,1}(t)$. Data is shown for 
the spin-$1/2$ Heisenberg chain with $L = 25$, where the
1D system is realized as a snake-like path through the lattice. Panel (e) shows 
a cut at $\ell = 1$. Even for shallow ${\cal R}$ with $d = 20$, results from 
$\ket{\psi_{{\cal R},\ell'}}$ are indistinguishable from a true Haar-random 
state. 
Dashed lines indicate power-law 
scalings of the correlations, cf.~Eq.~\eqref{Eq::Alpha_Beta}.} 
 \label{Fig1}
\end{figure}

{\it Introduction.}
Studying the properties of quantum many-body systems is 
tremendously challenging \cite{Feynman1982}. Notwithstanding significant 
progress  
thanks 
to
the development of sophisticated numerical methods \cite{schollwoeck20052011, 
weisse2006, Verstraete2008, Gull2011, Aoki2014, Carleo2017} 
and
groundbreaking experiments with 
cold-atom or trapped-ion platforms 
\cite{Bloch2012, Blatt2012}, simulations on universal quantum 
computers promise to yield major advancements in a multitude of research areas 
\cite{Georgescu2014, Tacchino2020}. While a fault-tolerant quantum 
computer is still far into the future, noisy 
intermediate-scale quantum (NISQ) devices are available and their 
current capabilities have been demonstrated for various problems such as 
electronic structure calculations \cite{Kandala2017, OMalley2016}, simulations 
of spectral functions \cite{Chiesa2019, Francis2020}, measurement of 
entanglement \cite{Choo2018, Wang2018}, topological phase transitions 
\cite{Smith2019_2}, and out-of-equilibrium dynamics~\cite{Lamm2018, Smith2019, 
Arute2020, Sommer2020}.

Recently, an important milestone towards so-called ``quantum supremacy'' 
\cite{Boixo2018} has been
achieved by using Google's NISQ device 
{\it Sycamore} \cite{Arute2019}. 
In the experiment, the Josephson junction based quantum processor was used to 
sample from the output distribution of \text{(pseudo-)}random circuits 
involving up to 53 
qubits, thereby going beyond the capacities of modern supercomputers.  
As this sampling task may appear rather abstract, it is crucial 
to 
identify a wider range of relevant applications of near-term NISQ devices 
which can be performed despite their imperfect 
fidelities of 
one- and two-qubit gates and 
the lack of error correction 
\cite{Preskill2018, Ippoliti2020, Gullans2020, Poggi2020}.   

Transport processes represent one of the most generic nonequilibrium 
situations \cite{Bertini2020}. In the quantum realm, the understanding of 
transport not only plays a key role to pave the way for future technologies 
such as spintronics \cite{Wolf2001}, but is also intimately related to 
fundamental questions of equilibration and thermalization in many-body systems 
\cite{dalessio2016, Borgonovi2016, Gogolin2016}. While quantum transport has 
been experimentally studied in 
mesoscopic systems, solid-state quantum magnets, and cold-atom 
settings (see e.g.\ \cite{DasSarma2011, Hess2019, Scheie2020, Hild2014,  
Jepsen2020}), active questions from the theory side include the quantitative 
calculation of transport coefficients \cite{Bertini2020, Rakovszky2020}, as 
well as explaining the emergence of conventional hydrodynamic transport from 
the underlying unitary time evolution of closed quantum systems 
\cite{Khemani2018}. 

In this Letter, we advocate near-term NISQ 
devices as useful platforms for simulating hydrodynamics in quantum many-body 
systems and, in particular, we 
show   
that random circuits (as realized in \cite{Arute2019})
form 
tailor-made building blocks for this purpose. With generalizations being 
possible \cite{DeRaedt2000} (see 
also
Supplemental Material \cite{SuppMat}), we specifically propose an 
efficient scheme to compute the infinite-temperature spatiotemporal
correlation function $C_{\ell,\ell'}(t)$ for one- and two-dimensional (1D, 2D) 
quantum 
spin systems,
\begin{equation}\label{Eq::Korrel}
 C_{\ell,\ell'}(t) = \text{Tr}[S_\ell^z(t) S_{\ell'}^z]/2^{L}\ , 
\end{equation}
where $S_{\ell(\ell')}^z$ is a spin-$1/2$ operator at lattice site $\ell$ 
($\ell'$), $S_\ell^z(t) = e^{i{\cal H}t} S_\ell e^{-i{\cal H}t}$ is the 
time-evolved operator with respect to (w.r.t.) some Hamiltonian ${\cal H}$, 
and $L$ 
denotes the number of spins (qubits). The spatiotemporal correlations 
$C_{\ell,\ell'}(t)$ are central objects for studying transport within linear 
response theory \cite{Bertini2020}, 
as well as thermalization and many-body localization in quantum  
systems \cite{Luitz2017}.
As a key ingredient, our scheme exploits 
the concept of quantum typicality 
\cite{Popescu2006, Goldstein2006, Reimann2007}, which asserts that 
ensemble averages can be accurately approximated by 
an expectation value w.r.t.\ a single pure state drawn 
at random from a high-dimensional Hilbert space \cite{Gemmer2004, lloydPhd, 
bartsch2009}. 
Remarkably, typicality 
applies independent 
of 
concepts such as the eigenstate thermalization 
hypothesis \cite{dalessio2016} and remains 
valid also 
for integrable or many-body localized systems~\cite{Heitmann2020}.

While random pure states have a long history for efficient numerical 
simulations \cite{Heitmann2020, Hams2000, iitaka2003, Alvarez2008, 
elsayed2013, monnai2014, steinigeweg2014, Richter2019,  Jin2020, 
Richter2019_2}, 
we 
demonstrate in this Letter that typicality can be used to recast the 
correlation function 
$C_{\ell,\ell'}(t)$
into a form which can be readily evaluated
on a quantum computer (see \cite{SuppMat} for a derivation),
\begin{equation}\label{Eq::Korrel_2}
 C_{\ell,\ell'}(t) = \frac{1}{2}\bra{\psi_{{\cal 
R},\ell'}(t)}S_\ell^z\ket{\psi_{{\cal 
R},\ell'}(t)} + {\cal O}(2^{-L/2})\ 
, 
\end{equation}
where 
 $\ket{\psi_{{\cal 
R},\ell'}(t)} = e^{-i{\cal H}t}\ket{\psi_{{\cal R},\ell'}}$, and 
$\ket{\psi_{{\cal 
R},\ell'}} = \ket{0}_{\ell'} \otimes {\cal R} \ket{0}^{\otimes L-1}$ results 
from the 
application 
of a \text{(pseudo-)}random circuit ${\cal R}$ on all qubits of the system 
except for the 
fixed reference site $\ell'$.
Importantly, as indicated by the second term on the right-hand side (r.h.s.), 
the 
accuracy of Eq.~\eqref{Eq::Korrel_2} improves exponentially with the 
size of the system \cite{Jin2020}.
Complementary to well-known approaches to obtain 
correlation 
functions such as Eq.~\eqref{Eq::Korrel} on a quantum computer 
\cite{Terhal2020, Somma2002, Pedernales2014} 
(see also \cite{Baez2020}), the  scheme 
proposed in this Letter operates without 
requiring an overhead of bath or ancilla qubits for initial-state preparation 
and measurement. Rather, it combines the random-circuit 
technology already 
realized on NISQ devices \cite{Arute2019} with ``quantum parallelism'' 
\cite{Alvarez2008, Schliemann2002} as the time-evolution of a single random 
state 
$\ket{\psi_{{\cal R},\ell'}}$ suffices to capture the full ensemble average 
\eqref{Eq::Korrel}. 
Furthermore, we particularly scrutinize the impact of Trotter and gate errors 
present in any realistic 
implementation and discuss the possibility to extract transport coefficients 
with error rates achievable on near-term hardware.
\begin{figure}[tb]
 \centering
 \includegraphics[width=0.95\columnwidth]{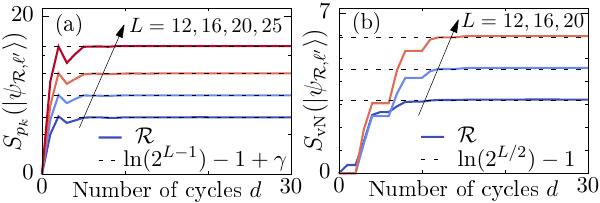}
 \caption{Buildup of randomness of $\ket{\psi_{{\cal R},\ell'}}$. (a) 
$S_{p_k}(\ket{\psi_{{\cal R},\ell'}})$ reaches the random-state value 
$\ln(2^{L-1}) - 1 
+ 
\gamma$ with Euler 
constant $\gamma \approx 0.577$ already at moderate $d$ \cite{Boixo2018}. (b)  
$S_\text{vN}(\ket{\psi_{{\cal R},\ell'}})$ approaches the ``Page value'' 
$\ln(2^{L/2})-1$ \cite{Page1993} appropriate for a random state on $L-1$ sites. 
The 
displayed $L$ values correspond to 
the 2D geometries $4\times 3$, $4\times 4$, $5 \times 4$, and $5 \times 5$. 
Data is averaged over $100$ realizations of ${\cal R}$.}
 \label{Fig2}
\end{figure}

{\it Description of Setup.} 
First, all qubits are initialized in the $\ket{0}$ state.
The algorithm then consists of a random circuit ${\cal R}$ acting on $L-1$ 
qubits followed by a
time evolution 
$U(t)$ on all $L$ sites [Fig.~\ref{Fig1}~(a)]. ${\cal R}$ comprises 
individual cycles, each composed of a 
layer of one-qubit gates and a layer of two-qubit gates, with $d$ denoting the 
total number of cycles [Fig.~\ref{Fig1}~(b)]. In each cycle, the one-qubit 
gates are randomly chosen from the set $\lbrace 
X^{1/2},Y^{1/2},T\rbrace$, 
where $X^{1/2}$ ($Y^{1/2}$) are $\pi/2$ rotations around the x-axis (y-axis) of 
the Bloch sphere and $T$ is the non-Clifford gate $T = \text{diag}(1,e^ 
{i\pi/4})$. We impose the constraint that the one-qubit gates on a given 
site have to be different in two subsequent cycles. As a two-qubit gate, 
we consider the controlled-Z (CZ) gate, 
CZ = $\text{diag}(1,1,1-1)$. (See \cite{SuppMat} for circuits with CNOT gates.) 
In each cycle, the CZ gates are aligned in one of 
the patterns ${\bf A}$-${\bf D}$ on a 2D geometry [Fig.~\ref{Fig1}~(c)], 
where we 
repeat the sequence ${\bf ABCD\dots}$ throughout ${\cal R}$, similar to Refs.\ 
\cite{Boixo2018, Arute2019}.  
After $d$ cycles, the state $\ket{\psi_{{\cal R},\ell'}} = \sum_k c_k 
\ket{k}$ is a superposition of computational basis states.  
It is the important realization that states generated 
from (shallow) random circuits ${\cal R}$ can approximate the properties of a 
Haar-random state \cite{Emerson2003, Oliveira2007, Boixo2018}, i.e., the 
coefficients $c_k$ are expected to closely follow a Gaussian distribution with 
zero mean. (Note that the exact preparation of a Haar-random state would be 
extremely inefficient in contrast \cite{Poulin2011}.)

For the subsequent time evolution, we exemplarily consider the 1D and 2D 
spin-$1/2$ Heisenberg model with 
nearest-neighbor interactions (see \cite{SuppMat} for results on another 
model \cite{Steinigeweg2014_2}), where we identify 
$\ket{0} = \ket{\uparrow}$ and $\ket{1} = 
\ket{\downarrow}$,   
\begin{equation}
 {\cal H} = \sum_{\langle \ell,\ell'\rangle} h_{\ell,\ell'}= \sum_{\langle 
\ell,\ell'\rangle} {\bf S}_\ell 
\cdot {\bf S}_{\ell'}\ ,\quad {\bf S}_\ell = (S_\ell^x, S_\ell^y, S_\ell^z)\ , 
\end{equation} 
where the 1D model is realized as a path through the 2D 
lattice [Fig.~\ref{Fig1}~(e)]. Focusing (for now) on 1D, the time-evolution 
operator 
$U(t) = \exp(-i{\cal H}t)$ is trotterized~\cite{DeVries1993}, 
\begin{equation}\label{Eq::Trotter}
 U(t) = \left(e^{-i{\cal H}\delta t}\right)^{N} \approx
\left(e^{-i{\cal H}_\text{e}\delta t}e^{-i{\cal H}_\text{o}\delta 
t}\right)^{N} + {\cal O}(\delta t^2)\ , 
\end{equation}
where ${\cal H}_\text{e}$ 
(${\cal H}_\text{o})$ denotes the even (odd) bonds $h_{\ell,\ell'}$ of ${\cal 
H}$, and $\delta t = t/N$ is a discrete time step. The mutually-commuting 
two-site terms
$\exp(-ih_{\ell,\ell'}\delta t)$ are then translated into elementary 
one- and two-qubit 
gates \cite{Tacchino2020} (We here use a representation which 
requires 
three CNOT gates \cite{Smith2019, SuppMat, Vatan2004}.)
Eventually, according to quantum 
typicality and our construction (see 
\cite{SuppMat}), a $z$-basis 
measurement of the qubit at site $\ell$ after time $t$ then yields the 
correlation function 
$2C_{\ell,\ell'}(t)$ [Figs.~\ref{Fig1}~(d),(e)].  
In particular, we show below that the correct extraction of 
$C_{\ell,\ell'}(t)$ remains possible even in the presence of inevitable 
Trotter and gate errors.

{\it Buildup of randomness.} 
In Fig.~\ref{Fig2}~(a), we study the growth of $S_{p_k}(\ket{\psi_{{\cal 
R},\ell'}}) 
= -
\sum_{k =1}^{2^L} p_k \ln p_k$ with $p_k = 
|c_k|^2$, which measures the spreading of $\ket{\psi_{{\cal R},\ell'}}$ within 
the 
computational basis due to ${\cal R}$. 
Moreover, the corresponding entanglement 
of 
$\ket{\psi_{{\cal 
R},\ell'}}$ is analyzed in Fig.~\ref{Fig2}~(b) by means of the von Neumann 
entropy $
 S_\text{vN}(\ket{\psi_{{\cal R},\ell'}}) 
= -\text{Tr}[\rho_A \ln \rho_A]$,
with $\rho_A = \text{Tr}_B\ket{\psi_{{\cal 
R},\ell'}}\bra{\psi_{{\cal R},\ell'}}$ being the reduced density matrix for a 
half-system bipartition into regions $A$ and $B$. Importantly, we observe that 
both $S_{p_k}$ and $S_\text{vN}$ reach their theoretically expected values for 
a 
random state \cite{Boixo2018, Page1993} already at moderate numbers of cycles 
$d \lesssim 10$, where the 
required $d$ appears to exhibit only a minor dependence on $L$ 
\cite{Boixo2018}.
We thus expect that $\ket{\psi_{{\cal R},\ell'}}$ 
mimics a true Haar-random state even for 
shallow ${\cal R}$ and can be 
used within the typicality approach to obtain $C_{\ell,\ell'}(t)$.
Throughout this Letter, we use a
fixed value $d = 20$, which yields very accurate results, see 
Fig.~\ref{Fig1}~(e) and \cite{SuppMat}. (Note that $d 
= 20$ has already been realized for 
$53$ qubits \cite{Arute2019}.) 
Eventually, we stress that this accuracy is achieved even though our design 
of ${\cal 
R}$ is not optimized \cite{Znidaric2007, Weinstein2008}, 
i.e., no particular 
fine-tuning of ${\cal R}$ appears to be necessary.
\begin{figure}[tb]
 \centering
 \includegraphics[width = 0.8\columnwidth]{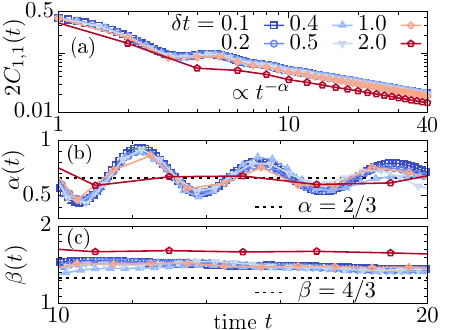}
 \caption{Impact of the Trotter time step. (a) $C_{1,1}(t)$ for varying values 
of $\delta t$. 
[(b),(c)] Extracted power-law exponents 
$\alpha(t)$ and $\beta(t)$. The dashed lines 
indicate the KPZ scaling~\cite{Bertini2020, Gopalakrishnan2019, 
Ljubotina2017, Ljubotina2019, Gopalakrishnan2019_2, DeNardis2019, Weiner2020}. 
Data is 
obtained 
for $L = 
25$ and $d = 20$.}
\label{Fig3}
\end{figure}

{\it Dependence on Trotter time step.}
Given the eponymous noise of NISQ devices, it is 
desirable to use as few gates as possible, i.e., a large time 
step $\delta t$. However, for a larger $\delta t$, the systematic 
error of the Trotter decomposition is in turn expected to increase [see 
r.h.s.\ of Eq.~\eqref{Eq::Trotter}]. In Fig.~\ref{Fig3}, we demonstrate that 
this expectation does not need to hold in practice (see Refs.\ \cite{Heyl2019, 
Sieberer2019}), such that a 
favorable trade-off between large $\delta t$ and acceptable Trotter error can 
be 
achieved.
Specifically, we find that  
the equal-site correlation function $C_{1,1}(t)$ in Fig.~\ref{Fig3}~(a) remains 
almost
unchanged for varying $\delta t$ between $\delta t = 0.1$ and $\delta t = 1$. 
Even though small deviations appear for larger
$\delta t = 2$, 
the qualitative shape of $C_{1,1}(t)$ remains similar also in 
this case. For a more detailed analysis, we consider the emerging hydrodynamic 
scaling 
of $C_{\ell,\ell'}(t)$ caused by the conservation of magnetization, 
$[{\cal H},\sum_\ell S_\ell^z] = 0$.
In particular, $C_{1,1}(t) \propto 
t^{-\alpha}$ develops a power-law tail for times $t \gtrsim 
10$ [Fig.~\ref{Fig3}~(a)], while 
correlations $C_{\ell,1}(t)$ build up 
throughout the system [cf.\ 
Fig.~\ref{Fig1}~(d)], i.e., $\Sigma^2(t) \propto t^{\beta}$ with the spatial 
variance 
\begin{equation}
 \Sigma^2(t) = \sum_\ell \ell^2 \widetilde{C}_{\ell,1}(t) - \big[\sum_\ell \ell 
\widetilde{C}_{\ell,1}(t)\big]^2\ , 
\end{equation}
where $\widetilde{C}_{\ell,1}(t) = C_{\ell,1}(t)/\sum_{\ell=1}^L C_{\ell,1}(t)$ 
with 
$\sum_\ell 
\widetilde{C}_{\ell,1}(t) = 1$.
In Figs.~\ref{Fig3}~(b) and \ref{Fig3}~(c), the impact of the Trotter step  
$\delta t$ on the instantaneous power-law 
exponents $\alpha(t)$ and $\beta(t)$ is studied for times $10 \leq t \leq 20$,
\begin{equation}\label{Eq::Alpha_Beta}
 \alpha(t) = -\frac{d \ln C_{1,1}(t)}{d \ln t}\ , \quad \beta(t) =\frac{d \ln 
\Sigma^2(t)}{d \ln t}\ .   
\end{equation}
We find that $\alpha(t)$ exhibits damped oscillations (presumably caused by the 
integrability of ${\cal H}$ \cite{Gopalakrishnan2019}) around the mean value 
$\alpha \approx 
2/3$, which signals superdiffusion and is  
consistent with a description of spin transport in terms of the 
Kardar-Parisi-Zhang (KPZ) 
universality class for the integrable and isotropic Heisenberg chain 
\cite{Bertini2020, Gopalakrishnan2019, 
Ljubotina2017, Ljubotina2019, Gopalakrishnan2019_2, DeNardis2019, Weiner2020}. 
Remarkably, 
$\alpha(t)$ is essentially independent 
of $\delta t$ and $\alpha \approx 2/3$ can be readily extracted even for the 
largest $\delta t = 2$.   
Likewise, $\beta(t)$ is found to remain stable up to $\delta t \leq 1$, 
albeit visible deviations now appear for $\delta t= 2$, which is 
explainable by the fact that $\beta(t)$ depends on the accuracy of the Trotter 
decomposition on the full system while $\alpha(t)$ is a local probe.  
Overall, the robustness of $C_{\ell,\ell'}(t)$ w.r.t.\ $\delta t$ 
is an important 
result and opens the door to reach 
nontrivial time scales with a manageable number of gates. 
For instance, fixing $\delta t = 1$, an evolution of $L = 25$ qubits up to $t 
= 20$ requires $2400$ one-qubit and $1440$ two-qubit gates in our case 
\cite{SuppMat}.
\begin{figure}[tb]
 \centering
 \includegraphics[width = 0.8\columnwidth]{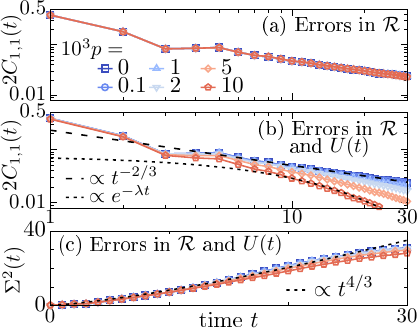}
 \caption{Impact of errors on [(a),(b)] $C_{1,1}(t)$ and (c)
$\Sigma^2(t)$. In (a) errors are present only within ${\cal R}$. Data is 
obtained for $L = 20$ and $d = 20$, averaged over $N_{\bf{r}} = 4500$ 
trajectories for a fixed design of ${\cal R}$ \cite{SuppMat}.}
 \label{Fig4}
\end{figure}

{\it Impact of noise.}
To model the impact of erroneous gates, we consider a
depolarization model with quantum 
channels ${\cal E}_\ell$ (${\cal E}_{\ell,\ell'}$) being applied after each 
one-qubit (two-qubit) gate \cite{Ippoliti2020}, 
\begin{align}
 {\cal E}_\ell(\rho) &= (1-p_1)\rho + \frac{p_1}{3}\sum_{\mu\neq 0} 
\sigma_\ell^\mu\rho \sigma_\ell^\mu\ , \label{Eq::Channel1}\\
 {\cal E}_{\ell,\ell'}(\rho) &= (1-p_2)\rho + 
\frac{p_2}{15}\sum_{(\mu,\nu)\neq (0,0)} 
\sigma_\ell^\mu \sigma_{\ell'}^\nu\rho \sigma_\ell^\mu \sigma_{\ell'}^\nu\ 
,\label{Eq::Channel2} 
\end{align}
where $\rho$ is the system's density matrix, $\sigma_\ell^\mu$ with $\mu = 
1,2,3$ are Pauli matrices, 
$\sigma_\ell^0 = \mathbb{1}$, and $p_1$ ($p_2$) are the one-qubit (two-qubit) 
error rates. We evaluate Eqs.~\eqref{Eq::Channel1} and 
\eqref{Eq::Channel2} by 
averaging over quantum trajectories \cite{Ippoliti2020, Dalibard1992}, $\rho(t) 
\approx (1/N_{\bf{r}}) 
\sum_{\bf{r}} \ket{\psi_{{\cal R},\ell'}^{\bf{r}}(t)} 
\bra{\psi_{{\cal 
R},\ell'}^{\bf{r}}(t)}$, where each trajectory $\ket{\psi_{{\cal 
R},\ell'}^{\bf{r}}(t)}$ 
corresponds to 
a particular history of one- and two-qubit Pauli errors.
In Figs.~\ref{Fig4}, we analyze the dynamics of $C_{\ell,\ell'}(t)$ 
obtained for a fixed time step 
$\delta t = 
1$ and varying error rates $p = p_2 = 10p_1$. First, we consider 
errors only within ${\cal R}$ and find that they have no effect on the 
equal-site correlator $C_{1,1}(t)$ [Fig.~\ref{Fig4}~(a)]. This 
exemplifies that typicality can also hold for states 
$\ket{\psi_{{\cal R},\ell'}}$ with non-Gaussian distributions of the 
coefficients 
$c_k$ in the computational basis \cite{Jin2020}. Specifically, the 
distribution of $p_k = |c_k|^2$ drifts from exponential to uniform for large 
error rates \cite{Boixo2018, SuppMat}. While this has been problematic for the 
sampling task in \cite{Arute2019}, it is irrelevant for our approach. 

In contrast, if errors are present in both ${\cal R}$ and $U(t)$ 
[Fig.~\ref{Fig4}~(b)], the decay of $C_{1,1}(t)$ depends on $p$. While a 
power-law tail $C_{1,1}(t) \propto t^{-\alpha}$ with $\alpha = 2/3$ can still 
be extracted for $p \lesssim 2\times 10^{-3}$ (roughly one order of magnitude 
smaller than currently achievable 
\cite{Arute2019}), the depolarization errors  
cause $C_{\ell,\ell'}(t)$ to decay exponentially for larger $p$ 
\cite{Ippoliti2020}.
Compared to the local probe $C_{1,1}(t)$,  
the spatial variance $\Sigma^2(t)$ appears 
to be less sensitive to noise, see Fig.~\ref{Fig4}~(c), and exhibits a 
power-law 
growth even for $p = 10^{-2}$. 
The robustness of $\Sigma^2(t)$ 
might be explained by the fact that $\ket{\psi_{{\cal R},\ell'}}$ is random 
and structureless at short times except for sites close to $\ell'$. 
Thus, errors away from $\ell'$ do not drastically alter the spreading 
of 
$C_{\ell,\ell'}(t)$ and the growth of $\Sigma^2(t)$. This is 
another  
result of this Letter. Given the robustness of $\Sigma^2(t)$ [and 
$C_{\ell,\ell'}(t)$] against Trotter and gate errors as well as the gradual 
improvement of technology, we expect 
near-term NISQ devices to provide a useful  platform to extract transport 
coefficients,  such as diffusion constants, of many-body quantum systems. In 
this context, the signal-to-noise ratio of the data in actual experiments can 
be systematically improved by increasing the number of repetitions 
\cite{Arute2019, Ippoliti2020, SuppMat}. 
\begin{figure}[tb]
 \centering
 \includegraphics[width = 0.9\columnwidth]{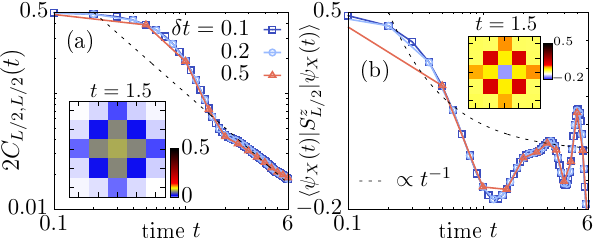}
 \caption{Dynamics in 2D. (a) 
$C_{L/2,L/2}(t)$ for varying $\delta t$. Dashed line indicates diffusive decay 
$\propto t^{-1}$. Inset shows $C_{\ell,L/2}(t)$ at $t 
= 1.5$. (b) The nonrandom state $\ket{\psi_X(t)}$ (see 
text for details) yields dynamics incompatible with diffusion.} 
 \label{Fig5}
\end{figure}

{\it Dynamics of 2D systems.}
Our approach is neither restricted to the dynamics of 1D systems 
nor to the choice of $\ell' = 1$. In Fig.~\ref{Fig5}~(a), we 
repeat our analysis of the $\delta t$ dependence for a 
2D Heisenberg model with $L = L_x \times 
L_y = 25$ and choose the reference site $\ell' = L/2$ as the central site of 
the lattice.
Analogous to the 1D case, we find that $C_{L/2,L/2}(t)$ is 
remarkably 
robust w.r.t.\ $\delta t$, with a stable hydrodynamic tail $C_{L/2,L/2}(t) 
\propto 
t^{-1}$, which signals the onset of conventional diffusion in 2D consistent 
with the transition from integrability to 
nonintegrability of 
${\cal H}$ from 1D to 2D \cite{Bertini2020}.  
Finally, let us consider the state 
$\ket{\psi_X} = \ket{\rightarrow}^{\otimes L/2-1} \otimes \ket{\uparrow} 
\otimes \ket{\rightarrow}^{\otimes L/2}$, 
i.e., a nonrandom product state where spins at $\ell \neq \ell'$
point in the $x$ direction, preparable by applying Hadamard gates on all but 
the central site. At 
$t = 0$, this state yields $\bra{\psi_X}S_{\ell}^z 
\ket{\psi_X} = 0.5 \delta_{\ell,L/2}$, i.e., the same as 
$\ket{\psi_{{\cal R},\ell'}}$.
The dynamics for $t > 0$ [Fig.~\ref{Fig5}~(b)], however, clearly differs from 
$C_{L/2,L/2}(t)$ and is incompatible with a 
power-law decay. Thus, the randomness of $\ket{\psi_{{\cal 
R},\ell'}}$ is crucial to extract the correct hydrodynamic scaling. This is 
another 
important result.

{\it Conclusion.}
We have shown that NISQ devices provide useful platforms to simulate 
hydrodynamics of quantum many-body systems. Relying on random-circuit 
technology and ``quantum parallelism'', we specifically presented an efficient 
scheme to obtain spatiotemporal 
correlation functions without the need of bath or ancilla qubits. As the 
intrinsic accuracy of
Eq.~\eqref{Eq::Korrel_2} improves exponentially with the number of qubits, we 
expect it to be scalable to 
larger systems. Especially for quantum many-body dynamics in 2D, which is known 
to be notoriously challenging for numerical methods, simulations on NISQ 
devices might help to answer open questions such as the existence of many-body 
localization.

Recently, ergodic and nonergodic behaviors have been shown in dual-unitary 
circuits \cite{Bertini2019, Bertini2019_2, Claeys2021}.
In a related work \cite{Claeys2021}, Claeys and Lamacraft 
also consider spatiotemporal correlations such as $C_{\ell,\ell'}(t)$. While 
Ref.\ \cite{Claeys2021} explores their dynamics for 
different classes of 
dual-unitary circuits, our work studies $C_{\ell,\ell'}(t)$ for 
explicit spin systems and, moreover, highlights the usefulness of random 
circuits for the preparation of suitable initial states. The role of typicality 
in dual-unitary circuits is a question for future work.

A natural extension would be to consider  thermal expectation values 
$\langle \bullet 
\rangle_\beta = \text{Tr}[\bullet e^{-\beta {\cal H}}]/\text{Tr}[e^{-\beta 
{\cal H}}]$ at inverse temperature 
$\beta$, which by virtue of typicality can be written as   
$\langle \bullet 
\rangle_\beta 
\approx \bra{\psi_\beta}\bullet\ket{\psi_\beta}/\braket{\psi_\beta|\psi_\beta}$ 
with 
$\ket{\psi_\beta} = e^{-\beta {\cal H}/2}\ket{r}$ 
\cite{sugiura2013}, where $\ket{r}$ is a random state. While 
$\ket{\psi_\beta}$ is straightforward to compute on a classical 
machine, a scheme to implement the unnatural nonunitary evolution on a quantum 
computer has been recently proposed \cite{Motta2020}. Thus, random circuits 
might also provide 
a means to prepare thermal states on NISQ devices, complementary to other 
approaches for this task~\cite{Motta2020, Temme2011, Cohn2020, 
Lu2020}. 

{\it Acknowledgements.} We sincerely thank F.\ Barratt, J.\ Dborin, H.\ De 
Raedt, A.\ G.\ Green, F.\ Jin, and R.\ 
Steinigeweg for helpful discussions and comments. This work was funded by the 
European Research 
Council 
(ERC) under the European Union's Horizon 2020 research and innovation programme
(Grant agreement No.\ 853368).

\clearpage
\newpage
%\onecolumngrid

% %%%%%%%%%%%%%%%%%%%%%%%%%
% % Supplemental Material %
% %%%%%%%%%%%%%%%%%%%%%%%%%
 
\setcounter{figure}{0}
\setcounter{equation}{0}
\renewcommand*{\citenumfont}[1]{S#1}
\renewcommand*{\bibnumfmt}[1]{[S#1]}
\renewcommand{\thefigure}{S\arabic{figure}}
\renewcommand{\theequation}{S\arabic{equation}}

\section*{Supplemental material}

\subsection{Derivation of Eq.\ (2)}

Let us show how typicality can be used to recast the 
correlation function $C_{\ell,\ell'}(t)$ from Eq.\ (1) into 
a form which can be readily 
evaluated on a quantum computer. 
We begin by rewriting Eq.~(1) as  
\begin{align}
 C_{\ell,\ell'}(t) &= \text{Tr}[S_\ell^z(t) 
P_{\ell'}^\uparrow]/2^L - \text{Tr}[S_\ell(t)]/2^{L+1} \label{Eq::D1}\\
&=\text{Tr}[P_{\ell'}^\uparrow S_\ell^z(t) 
P_{\ell'}^\uparrow]/2^L\ , \label{Eq::D2} 
\end{align}
where $P_{\ell'}^\uparrow = S_{\ell'}^z+1/2 = 
({\cal P}_{\ell'}^\uparrow)^2$ is a projection onto the $\ket{\uparrow}$ state 
of the spin at site $\ell'$. Moreover, from Eq.~\eqref{Eq::D1} to 
Eq.~\eqref{Eq::D2}, we have used the cyclic invariance 
of the trace and $\text{Tr}[S_\ell^z] = 0$. 
Let now $\ket{r}$ be a pure state drawn at 
random according to the unitary invariant Haar measure,
\begin{equation}
 \ket{r} = \sum_{k=1}^{2^L} c_k \ket{k}\ , 
\end{equation}
i.e., 
the real and 
imaginary parts of the $c_k$ are Gaussian random numbers with zero 
mean (constrained by $\sum_k |c_k|^2 = 1$) with $\ket{k}$ denoting the 
orthogonal computational basis states. According to typicality, the trace 
$\text{Tr}[\cdot]/2^L$ can then be approximated as~\cite{Jin2020S}    
\begin{equation}\label{Eq::Typicality}
 \text{Tr}[P_{\ell'}^\uparrow S_\ell^z(t) 
P_{\ell'}^\uparrow]/2^L = \bra{r} P_{\ell'}^\uparrow S_\ell^z(t) 
P_{\ell'}^\uparrow\ket{r} + {\cal O}(2^{-L/2})\ ,  
\end{equation}
where the second term on the right hand side indicates that the 
statistical error vanishes exponentially with the 
size of the system \cite{Jin2020S} (and can often be neglected already 
for 
moderate values of $L$ \cite{Richter2019_2S, 
Jin2020S}). 
Defining now 
\begin{equation}
 \ket{\psi_{\text{Haar},\ell'}} = P_{\ell'}^\uparrow 
\ket{r}/||P_{\ell'}^\uparrow \ket{r}||\ , 
\end{equation}
with $||P_{\ell'}^\uparrow \ket{r}||^2 
\approx (1/2) \braket{r|r} = 1/2$, and interpreting the time dependence as a 
property of the state, it follows from 
Eq.~\eqref{Eq::Typicality} that 
\begin{equation}
 C_{\ell,\ell'}(t) \approx (1/2)\bra{\psi_{\text{Haar},\ell'}(t)} S_\ell^z 
\ket{\psi_{\text{Haar},\ell'}(t)}, 
\end{equation}
which is formally equivalent to 
Eq.~(2) upon identifying
$\ket{\psi_{{\cal R},\ell'}} \leftrightarrow \ket{\psi_{\text{Haar},\ell'}}$. 
On current NISQ 
devices, the state $\ket{\psi_{{\cal R},\ell'}}$ with approximately Haar-random 
coefficients can be efficiently generated by a \text{(pseudo-)}random circuit 
${\cal R}$ \cite{Arute2019S, Boixo2018S}. 
Furthermore, while Gaussian coefficients $c_k$ are  
preferential as their distribution $P(c_k)$  
then remains Gaussian also in the eigenbasis of ${\cal H}$, the exact 
distribution (given enough randomness) often turns out to be unimportant  
for 
the applicability of 
typicality \cite{Jin2020S}. We have demonstrated this fact in the 
context of Fig.~4~(a), where $C_{\ell,\ell'}(t)$ was shown to be 
robust against depolarization errors within ${\cal R}$. 

\subsection{Accuracy of the typicality approximation on a quantum computer}

\begin{figure}[tb]
 \centering
\includegraphics[width = 0.85\columnwidth]{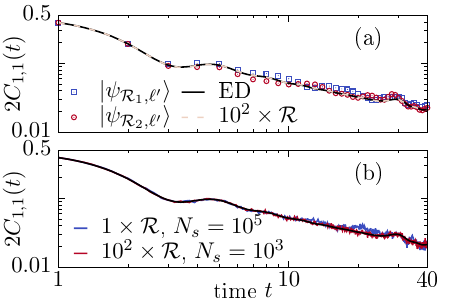}
\caption{Accuracy of the typicality approach in comparison with exact 
diagonalization for a 1D Heisenberg chain with $L = 16$. (a) Results 
obtained from two different realizations of ${\cal R}$ show visible 
fluctuations, while an averaging over $100$ instances of ${\cal R}$ yields 
results indistinguishable from ED. (b) The averaging over ${\cal R}$ can be 
combined with the necessary sampling over circuit outputs to reconstruct 
$C_{1,1}(t)$ in an actual 
experiment. $N_s$ here denotes the number of samples for each individual ${\cal 
R}$. Even though the total number of runs is the same, the data averaged over 
$100\times {\cal R}$ agrees better with ED.}
\label{FigS1}
\end{figure}

According to Eq.~(2), the 
infinite-temperature spatiotemporal correlation function 
$2C_{\ell,\ell'}(t)$ can be obtained as the expectation value $
\bra{\psi_{{\cal 
R},\ell'}(t)}S_\ell^z\ket{\psi_{{\cal R},\ell'}(t)}$. This relation is based on 
the concept of 
typicality, the accuracy of 
which improves exponentially with the size of the system. While this accuracy 
has been already demonstrated in Figs.\ 3 - 5, 
we provide further evidence in Fig.\ 
\ref{FigS1}~(a), where we compare results obtained from two different 
realizations of the random circuit ${\cal R}$ to exact diagonalization (ED) 
data for a system of size $L = 16$. 
Even for this rather small value of $L$, we find that the dynamics obtained 
from $\ket{\psi_{{\cal R}_1,\ell'}}$ and $\ket{\psi_{{\cal R}_2,\ell'}}$ 
closely follow 
the exact result, albeit some small fluctuations are visible at longer 
times. In this context, we note that the accuracy of typicality can be further 
improved by averaging over the output of 
different random states, i.e., over different realizations of ${\cal R}$. As 
shown in Fig.\ \ref{FigS1}~(a), averaging over $10^2$ realizations of 
${\cal R}$ yields results 
indistinguishable from ED. For larger systems such as $L = 25$ in Fig.\ 3, 
averaging is not necessary and a single random state is 
sufficient to 
yield negligibly small statistical errors.    

So far, we have focused directly on the expectation value 
$\bra{\psi_{{\cal 
R},\ell'}(t)}S_\ell^z\ket{\psi_{{\cal R},\ell'}(t)}$. This expectation value, 
however, can not 
be obtained on a quantum computer in a single run. Specifically, the 
measurement of the qubits $1 - L$ at the end of the algorithm merely yields a 
single state in the computational basis such as $\ket{0 1 0 1\dots}$ or 
$\ket{1111\dots}$, while $\ket{\psi_{{\cal R},\ell'}(t)}$ will in general be a 
superposition of all these states, 
\begin{equation}
 \ket{\psi_{{\cal R},\ell'}(t)} = \sum_{k=1}^{2^L} a_k \ket{k}\ . 
\end{equation}
The full expectation value $C_{\ell,\ell'}(t)$ can then be reconstructed by 
repeating the 
experiment multiple times, 
\begin{equation}\label{Eq::Reconstruct} 
2C_{\ell,\ell'}(t) = \frac{1}{2}\left(\sum_{\ket{k}, \ell = \uparrow} 
|\widetilde{a}_k|^2 - \sum_{\ket{k}, \ell = \downarrow} |\widetilde{a}_k|^2 
\right)\ , 
\end{equation}
where $|\widetilde{a}_k|^2$ is the experimentally obtained probability of the 
state $\ket{k}$, and the sums rum over all states $\ket{k}$ for which the 
spin $\ell$ is found to be up or down respectively. (Once again we identify 
$\ket{0} = \ket{\uparrow}$ and $\ket{1} = \ket{\downarrow}$.) By increasing the 
number of repetitions, the accuracy can be systematically improved, 
$|\widetilde{a}_k|^2 \to |a_k|^2$.
In Fig.~\ref{FigS1}~(b), we show that this sampling of the distribution of the 
$|a_k|^2$ can be combined with the averaging over different random circuits 
${\cal R}$ to yield accurate results. Specifically, we compare results from
one realization of ${\cal R}$ with $N_s = 10^5$ repetitions to data obtained 
from $100$ realizations of ${\cal R}$ with only $N_s = 10^3$ repetitions each, 
i.e., the total number of experimental runs is the same in both cases. While 
the noise of the data is very similar in both cases, the averaging over 
different ${\cal R}$ yields a better agreement with ED. 
We note that varying the design of ${\cal R}$ on a NISQ device should be 
straightforward experimentally. Moreover, the number of experimental runs used 
in Fig.~\ref{FigS1} would execute very quickly \cite{Arute2019S}.

\subsection{Impact of noise on probability distribution}

\begin{figure}[tb]
 \centering
\includegraphics[width = 0.8\columnwidth]{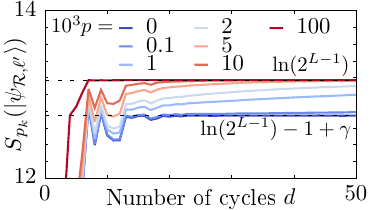}
\caption{Altering of the output probability distribution of the state 
$\ket{\psi_{\cal R}}$ due to depolarization errors. For increasing error rate 
$p$ or increasing number of cycles $d$, $S_{p_k}(\ket{\psi_{{\cal R},\ell'}})$ 
drifts 
from the Gaussian value $\ln(2^{L-1}) - 1 + \gamma$ towards the value 
$\ln(2^L-1)$ which corresponds to a uniform distribution of the $p_k$. Data is 
obtained by 
averaging over $N_{{\bf r}} = 5000$ trajectories for the random circuit shown 
in Fig.~\ref{FigS3}.}
\label{FigS2}
\end{figure}

In Fig.~4, we have shown that erroneous  gates within ${\cal R}$ turn 
out to be unimportant for the typicality approach presented in this Letter.
However, such errors do have an impact on the 
probability distribution $P(c_k)$ which characterizes the state 
$\ket{\psi_{{\cal R},\ell'}}$ \cite{Boixo2018S}. This is visualized in 
Fig.~\ref{FigS2} 
where we study the spreading of $\ket{\psi_{{\cal R},\ell'}}$ in the 
computational 
basis, analogous to Fig.~2~(a). 
Given the trajectory approach to unravel the quantum channels, 
$S_{p_k}(\ket{\psi_{{\cal R},\ell'}})$ is now defined as
\begin{equation}
 S_{p_k}(\ket{\psi_{{\cal R},\ell'}}) = \sum_{k=1}^{2^L} \rho_{kk}\ln 
\rho_{kk}\ ,
\end{equation}
where $\rho=(1/N_{{\bf r}})\sum_{{\bf r}}\ket{\psi_{{\cal R},\ell'}^{\bf{r}}} 
\bra{\psi_{{\cal 
R},\ell'}^{\bf{r}}}$ approximates the system's density matrix. We find that for 
increasing error rate $p$ (or increasing number of cycles $d$), 
$S_{p_k}(\ket{\psi_{{\cal R},\ell'}})$ exhibits a drift from $\ln(2^{L-1}) -1 + 
\gamma$ 
towards $\ln(2^{L-1})$. While the former corresponds to a Gaussian distribution 
of the $c_k$ (i.e., an exponential distribution of $p_k = |c_k|^2$), the latter 
signals a uniform distribution of the $p_k$.

\begin{figure}[b]
 \centering
\includegraphics[width = 0.85\columnwidth]{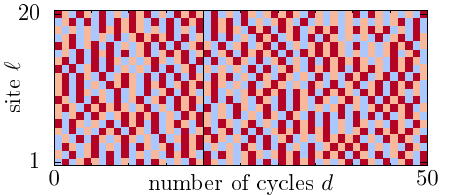}
\caption{Visualization of the random circuit ${\cal R}$ used to produce the 
data in Figs.~4 and \ref{FigS2}. Different colors signal different 
one-qubit gates from the set $\lbrace X^{1/2}, Y^{1/2}, T\rbrace$. The dynamics 
in Fig.~4 was computed for $d = 20$, as indicated by the black 
vertical line. }
\label{FigS3}
\end{figure}

We note that the averages over quantum trajectories in Figs.~4 and 
\ref{FigS2} have been obtained for a single realization of the random circuit 
${\cal R}$, i.e., a fixed sequence of one-qubit and two-qubit gates with 
errors being randomly interspersed in each run.  
This specific realization of ${\cal R}$ is visualized in Fig.~\ref{FigS3}. Note 
that due to the additional costs of averaging over trajectories, we have chosen 
a slightly smaller system with $L = L_x \times L_y = 5 \times 4 = 20$.

\subsection{Random circuit with CNOT gates}
\begin{figure}[tb]
 \centering
\includegraphics[width = 0.8\columnwidth]{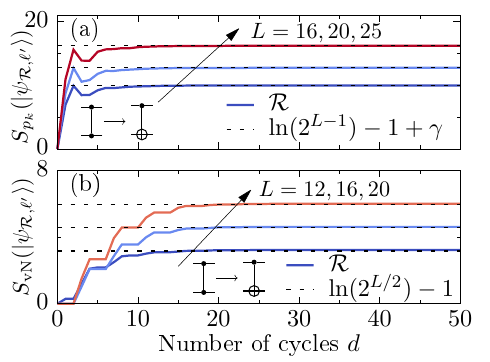}
\caption{Analogous data as in Fig.\ 2, but now the two-site 
gates within ${\cal R}$ are chosen as CNOT gates instead of CZ gates. Once 
again $S_{p_k}(\ket{\psi_{{\cal R},\ell'}})$ and $S_\text{vN}(\ket{\psi_{{\cal 
R},\ell'}})$ 
approach their expected values for a random state with increasing cycle number 
$d$.}
\label{FigS4}
\end{figure}

In Fig.~\ 2 of the main text, we have shown that the application of 
the \text{(pseudo-)}random circuit ${\cal R}$ yields a state $\ket{\psi_{{\cal 
R},\ell'}}$ which mimics a true Haar-random state, at least with respect to the 
quantities $S_{p_k}$ and $S_\text{vN}$.
In Fig.~\ref{FigS4}, we show that the behavior found in Fig.~2 is not 
caused by ${\cal R}$ being fine-tuned. Specifically, we find that 
$S_{p_k}(\ket{\psi_{{\cal 
R},\ell'}})$ and $S_\text{vN}(\ket{\psi_{{\cal 
R},\ell'}})$ quickly approach the expected values for a random state also if 
the CZ 
gates are replaced by CNOT gates. Comparing Figs.\ 2 and \ref{FigS4}, 
however, the convergence seems to be slightly faster in the former case. 

\subsection{Generalization of the typicality relations}

Relying on the concept of typicality, we show in this Letter that 
\text{(pseudo-)}random circuits ${\cal R}$ are useful building blocks to study 
quantum many-body systems.
The important realization is that states $\ket{{\cal R}} = {\cal R} 
\ket{0}^{\otimes L}$ generated from such a circuit can faithfully represent the 
properties of a true Haar-random state $\ket{r}$.
In the main text, we have exemplified this approach by considering 
infinite-temperature spatiotemporal correlation functions 
$C_{\ell,\ell'}(t)$ for one- and two-dimensional quantum spin systems. The 
overall scheme, however, can be applied in a more general context, which we 
outline below. (Obviously, instead of spin systems, one could likewise 
consider fermionic or bosonic models.) One additional application 
of random circuits would be the preparation of thermal states 
at finite temperature, which we have already mentioned in the main text. Here, 
however, we focus on simulations of general correlation functions and of the 
density of states.

\subsubsection{Correlation functions}

Let $A$ and $B$ denote two hermitian operators and ${\cal D}$ the dimension 
of the Hilbert space. Then the infinite-temperature correlation function
$\langle A(t) B \rangle_{\infty}$ is defined as
\begin{equation}
 \langle A(t) B \rangle_{\infty} = \frac{\text{Tr}[A(t) 
B]}{{\cal D}}\ . 
\end{equation}
Without loss of generality, we now assume that $\text{Tr}[A] = \text{Tr}[A(t)] 
= 0$. Then, $\langle A(t) B \rangle_{\infty}$ can be formally rewritten as 
\begin{align}
 \frac{\text{Tr}[A(t) 
B]}{{\cal D}} &=  \frac{\text{Tr}[A(t) 
(B+\epsilon)]}{{\cal D}} \\ &= \frac{\text{Tr}[\sqrt{B+\epsilon}A(t) 
\sqrt{B+\epsilon}]}{{\cal D}}\ , 
\end{align}
where $\epsilon$ is chosen such that the spectrum of $B + \epsilon$ is 
non-negative and the square-root operation has to be understood in the 
eigenbasis of $B$.
Exploiting typicality, we can approximate the trace by an expectation value 
with respect to a random state $\ket{{\cal R}}$ (generated by a 
random circuit), 
\begin{equation}\label{Eq::Deri}
\langle A(t) B \rangle_{\infty} = \frac{\bra{{\cal R}}\sqrt{B+\epsilon}A(t) 
\sqrt{B+\epsilon}\ket{{\cal R}}}{\braket{{\cal R}|{\cal R}}} + {\cal 
O}(\frac{1}{\sqrt{{\cal D}}})\ , 
\end{equation}
where the statistical error of this approximation vanishes with the inverse 
square-root of the Hilbert-space dimension. For an interacting system, ${\cal 
D}$ grows exponentially with the system size. From Eq.~\eqref{Eq::Deri}, it 
follows that  
\begin{equation}\label{Eq::Final}
 \langle A(t) B \rangle_{\infty} = c\bra{\widetilde{\cal 
R}(t)}A\ket{\widetilde{\cal R}(t)}\ , 
\end{equation}
where $\ket{\widetilde{\cal R}} = \sqrt{B+\epsilon}\ket{\cal 
R}/||\sqrt{B+\epsilon}\ket{\cal R}||$ and $c = ||\ket{\widetilde{\cal 
R}}||^2/||\ket{\cal R}||^2$.
Equation \eqref{Eq::Final} is a generalization of Eq.~(2) from 
the main text. We note that the construction of the state $\ket{\widetilde{\cal 
R}}$ can be difficult in practice as the application of the square-root in 
principle requires the diagonalization of $B$. Assuming that $B$ is a local 
operator which only acts nontrivially on a few qubits, a full diagonalization 
can be circumvented however, and an efficient preparation of 
$\ket{\widetilde{\cal 
R}}$ might remain possible \cite{Richter2019_2S}. A significant 
simplification can be achieved if $B+\epsilon = P$ is a projection, $P = P^2$. 
In this case, no diagonalization is required. Such a scenario applies to the 
spatiotemporal correlation function $C_{\ell,\ell'}(t)$ studied in this Letter. 
In particular, we have $B = S_{\ell'}^z$ and $\epsilon = 1/2$. 

\subsubsection{Density of states}

Here, we briefly outline an algorithm to obtain the density of states (DOS)
$\Omega(E)$ of some Hamiltonian ${\cal H}$ on a quantum computer by means of 
random states, which was first presented in \cite{DeRaedt2000S}.   
The DOS is defined as 
\begin{equation}\label{Eq::Omega}
 \Omega(E) = \sum_i \delta (E-E_i) = \frac{1}{2\pi} \int_{-\infty}^\infty 
e^{iEt} \text{Tr}[e^{-i{\cal H}t}]\ dt\ , 
\end{equation}
where we have used the definition of the $\delta$ function. Once again, the 
trace on the right hand side can be approximated as 
\begin{equation}
 \text{Tr}[e^{-i{\cal H}t}] \approx \bra{\cal R}e^{-i{\cal H}t}\ket{\cal R} = 
\braket{{\cal R}|{\cal R}(t)}\ . 
\end{equation}
Instead of evolving the state $\ket{{\cal R}(t)}$ in time and projecting the 
initial state $\ket{\cal R}$ onto it, it is helpful to realize that 
$\text{Tr}[e^{-i{\cal H}t}] = \text{Tr}[e^{-i{\cal H}t/2}e^{-i{\cal H}t/2}]$. 
Thus, we can write
\begin{equation}
 \text{Tr}[e^{-i{\cal H}t}] \approx \braket{{\cal R}(t/2)|{\cal R}(t/2)}\ , 
\end{equation}
where the accuracy of the approximation, analogous to Eq.~\eqref{Eq::Deri},
improves exponentially with the size of the system. The fact 
that the Fourier transform in Eq.~\eqref{Eq::Omega} can be carried out only up 
to a finite time leads to a broadening of the individual energy peaks. 
Increasing the time $t$ allows to obtain $\Omega(E)$ with a better and better 
resolution.

\subsection{Decomposition of spin-exchange terms into elementary gates}

\begin{figure}[h]
 \centering
\includegraphics[width = 0.9\columnwidth]{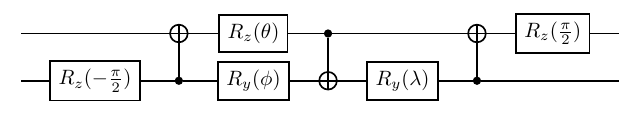}
\caption{Decomposition of the two-site operator $\exp(-ih_{\ell,\ell'}t)$ into 
elementary one- and two-qubit gates, where $h_{\ell,\ell'} = S_\ell^x 
S_{\ell'}^x + S_{\ell}^y S_{\ell'}^y + S_\ell^z S_{\ell'}^z$. The decomposition 
requires three CNOT gates and five one-qubit rotations. The angles of the 
one-qubit rotations are given by $\theta = \lambda = \pi/2 - 1/2$ and $\phi = 
1/2 - \pi/2$ \cite{Smith2019S}.}
\label{FigS5}
\end{figure}

There exist different possibilities to decompose the time-evolution operator 
$\exp(-ih_{\ell,\ell'}t)$ for a two-site  Heisenberg Hamiltonian into 
elementary one- and two-qubit gates. Since two-qubit gates typically have 
a larger error rate, we here use a representation which only requires three 
CNOT gates as well as five one-qubit rotations \cite{Vatan2004S, Smith2019S}, 
see Fig.\ \ref{FigS5} for details. 
A single step on $L = 25$ qubits 
(i.e., $24$ bond terms) would therefore require $5 \times 24 = 120$ one-qubit 
and $3 \times 24 = 72$ CNOT gates. Fixing $\delta t = 1$, a time evolution up 
to $t = 20$ thus involves $120 \times 20 = 2400$ one-qubit and $72\times 20 = 
1440$ CNOT gates. 

\subsection{Dependence of dynamics on depth of ${\cal R}$}

\begin{figure}[tb]
 \centering
 \includegraphics[width=0.95\columnwidth]{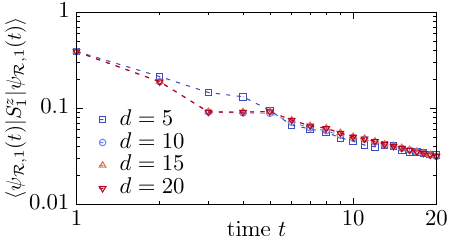}
 \caption{Expectation value $\bra{\psi_{{\cal R},1}(t)}S_1^z\ket{\psi_{{\cal 
R},1}(t)}$ for random circuits ${\cal R}$ with different depths $d = 
5,10,15,20$. Data is analogous to Fig.~1~(e) of the main text. For sufficiently 
large $d$, the expectation value converges towards the correlation function 
$2C_{1,1}(t)$. Data is shown for the 1D Heisenberg chain with system size $L 
= 25$. The Trotter time step is fixed to $\delta t = 0.5$.}
 \label{FigS6}
\end{figure}

In the main text, we have presented numerical results for a fixed depth $d = 
20$ of the random circuit ${\cal R}$. As exemplified in Fig.~1~(e), this depth 
turned out sufficient such that the correlation function $C_{1,1}(t)$ obtained 
from the state $\ket{\psi_{{\cal R},\ell'}}$ is indistinguishable from that of 
a true Haar-random state. In Fig.\ \ref{FigS6}, we now present additional 
results 
for shallower random circuits ${\cal R}$ with $d = 5,10,15$, for which the 
resulting state $\ket{\psi_{{\cal R},\ell'}}$ is consequently less random.
While the data 
for $d = 10$ and $d = 15$ agree very well with our previous results for $d = 
20$, deviations occur for the shallowest circuit with $d = 5$. However, even in 
the latter case, the emerging long-time hydrodynamic tail is very similar to 
the larger choices of $d$, albeit fluctuations are slightly more pronounced.  
Thus, even for moderately random states, for which the entanglement entropy 
differs from the Page value (cf.\ Fig.~2 from 
the main text), the resulting dynamics is still a good approximation to the 
autocorrelation function $C_{1,1}(t)$, and correctly captures the emerging 
hydrodynamic behavior.

We here leave it to future work to 
study the dependence of $\bra{\psi_{{\cal R},\ell'}(t)}S_\ell^z\ket{\psi_{{\cal 
R},\ell'}(t)}$ on the depth of ${\cal R}$ in more detail. In particular, it 
will be an interesting direction to analyze the impact of spatial variations of 
the randomness of $\ket{\psi_{{\cal 
R},\ell'}}$ on non-local correlations with 
$\ell 
\neq \ell'$. In this context, it might also be insightful to consider initial 
conditions where the system is split into patches of Haar-random 
states, with no initial entanglement between different patches 
\cite{Arute2019S}. 

\subsection{Extraction of diffusion constant for a nonintegrable spin-ladder 
model}

In the main text, we have restricted ourselves to the analysis of the power-law 
exponents $\alpha(t)$ and $\beta(t)$ which have indicated the emergence of 
superdiffusive (diffusive) transport in the 1D (2D) Heisenberg model. Let us 
now demonstrate that our scheme can also quantitatively capture the correct 
diffusion constant in the case of a nonintegrable model. To this end, we focus 
on the quasi-1D XY ladder,
\begin{align}\label{Eq::Ham_XYLadder}
 {\cal H} &= \sum_{\ell = 1}^{L_x-1} \sum_{k=1}^2 \left(S_{\ell,k}^x 
S_{\ell+1,k}^x+ S_{\ell,k}^y S_{\ell+1,k}^y\right) \\ &+ \sum_{\ell = 
1}^{L_x}\left(S_{\ell,1}^x 
S_{\ell,2}^x+ S_{\ell,1}^y S_{\ell,2}^y\right) \nonumber\ , 
\end{align}
where $L_x$ denotes the number of rungs. The high-temperature spin diffusion 
constant of the XY ladder is well-known to be $D\approx 
0.95$ \cite{Steinigeweg2014S}, and the model is a popular example to benchmark 
numerical methods for transport coefficients \cite{Rakovszky2020S}.

\begin{figure}[tb]
 \centering
 \includegraphics[width=1\columnwidth]{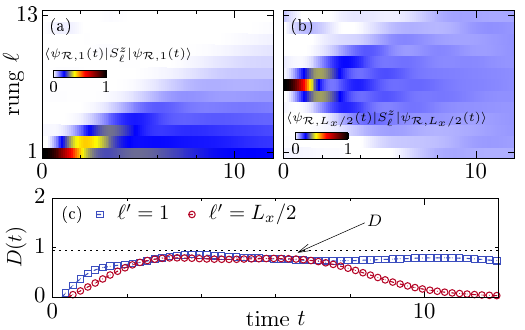}
 \caption{Expectation value $\bra{\psi_{{\cal R},\ell'}(t)}S_\ell^z 
\ket{\psi_{{\cal R},\ell'}(t)}$ for the XY ladder 
\eqref{Eq::Ham_XYLadder} where the reference rung $\ell'$ is chosen as (a) 
$\ell' = 1$ or (b) $\ell' = L_x/2$. (c) Time-dependent diffusion coefficient 
$D(t)$ extracted from the correlation profiles in (a) and (b). The diffusion 
constant $D$ is consistent with known results from 
the literature \cite{Rakovszky2020S, Steinigeweg2014S} (horizontal dashed 
line). Note that the drop of $D(t)$ in the case of $\ell = L_x/2$ for $t 
\gtrsim 8$ indicates the onset of finite-site effects. 
We have $L_x = 13$ and $\delta t = 0.1$ in all cases.}
 \label{FigS7}
\end{figure}

In Fig.\ \ref{FigS7}, we show numerical data obtained by the approach outlined 
in the main text. Specifically, we consider a grid of $13 \times 2$ qubits, 
i.e., $26$ qubits in total, and perform a (pseudo)random circuit on the ladder, 
except for the reference rung $\ell'$. We present two examples, namely, $\ell' 
= 1$ [edge of the ladder, see Fig.\ \ref{FigS7}~(a)] and $\ell' = L_x/2$ 
[center of the ladder, see Fig.\ \ref{FigS7}~(b)]. Subsequently, the resulting 
state $\ket{\psi_{{\cal R},\ell'}}$ is evolved in time with respect to ${\cal 
H}$ such that correlations spread throughout the system. The time-dependent 
diffusion coefficient $D(t)$ can then be extracted from the spatial variance 
$\Sigma^2(t)$ of the correlation profile [cf.~Eq.~(5) in the main text] 
according to $D(t) = 
\partial_t \Sigma^2(t)/2$ ($\ell' = L_x/2$), or $D(t) = \partial_t 
\Sigma^2(t)$ ($\ell' = 1$, as correlations only spreads in one direction in 
this case). 

The resulting data for $D(t)$ is 
shown in Fig.\ \ref{FigS7}~(c). Above a mean-free time $t 
\gtrsim 2$, we find that $D(t)$ becomes roughly 
time-independent, i.e., it becomes a genuine diffusion {\it constant}, 
$D(t) \to D$. In particular, $D$ is almost independent of the 
choice of $\ell'$ in the intermediate time-window $2 \lesssim t \lesssim 8$, 
and is consistent with results from other numerical approaches [dashed line in 
Fig.\ 
\ref{FigS7}~(c)]
\cite{Steinigeweg2014S, Rakovszky2020S}. In this context, let us stress that in 
some cases it is not 
advisable to initialize the density peak at the edges of the system, as edge 
effects might influence the dynamics. For the nonintegrable spin ladder 
considered here, however, the choices of $\ell' = 1$ or $\ell' = L_x/2$ yield 
consistent results. 

Eventually, for $t > 8$, we find that $D(t)$ starts to decrease again in the 
case of $\ell' 
= L_x/2$. This can be understood as a finite-size effect as the correlation 
profile reaches the boundaries of the ladder at these times, cf.\ Fig.\ 
\ref{FigS7}~(b). In constrast, finite-size effects are much less pronounced for 
$\ell' = 1$, such that $D(t)$ remains constant on longer time scales, which is 
beneficial for the extraction of $D$.

 \end{document}